 \documentclass[final,5p,times,twocolumn,authoryear]{elsarticle}

\usepackage{amssymb}

\usepackage{lineno,hyperref}
\usepackage{times}
\usepackage{graphicx}
\usepackage{epstopdf}
\usepackage{booktabs} 
\usepackage{subfigure}

\usepackage{natbib}
\biboptions{sort&compress}

\begin{document}
 
\begin{frontmatter}



\title{Improving GPU-accelerated Adaptive IDW Interpolation Algorithm \\Using 
	   Fast \textit{k}NN Search}


\author[myaddress]{Gang Mei}
\author[myaddress]{Nengxiong Xu\corref{mycorrespondingauthor}}
\cortext[mycorrespondingauthor]{Corresponding author}\ead{xunengxiong@cugb.edu.cn}
\author[myaddress]{Liangliang Xu}
\address[myaddress]{School of Engineering and Technology, China University of Geosciences, Beijing 100083, China}

\begin{abstract}
This paper presents an efficient parallel Adaptive Inverse Distance 
Weighting (AIDW) interpolation algorithm on modern Graphics Processing Unit 
(GPU). The presented algorithm is an improvement of our previous 
GPU-accelerated AIDW algorithm by adopting fast $k$-\textbf{N}earest 
\textbf{N}eighbors ($k$NN) search. In AIDW, it needs to find several nearest 
neighboring data points for each interpolated point to adaptively determine 
the power parameter; and then the desired prediction value of the interpolated point 
is obtained by weighted interpolating using the power parameter. In this 
work, we develop a fast $k$NN search approach based on the space-partitioning 
data structure, even grid, to improve the previous GPU-accelerated AIDW algorithm. 
The improved algorithm is composed of the stages of $k$NN search and weighted 
interpolating. To evaluate the performance of the improved algorithm, we 
perform five groups of experimental tests. Experimental results show that: 
(1) the improved algorithm can achieve a speedup of up to 1017 over the 
corresponding serial algorithm; (2) the improved algorithm is at least two 
times faster than our previous GPU-accelerated AIDW algorithm; and (3) the utilization 
of fast $k$NN search can significantly improve the computational efficiency of 
the entire GPU-accelerated AIDW algorithm.
\end{abstract}


\end{frontmatter}







\section{Introduction}
\label{sec1:introduction}

Spatial interpolation is a fundamental tool in Geographic Information System 
(GIS). The most frequently used spatial interpolation algorithms include the 
Inverse Distance Weighting (IDW) \citep{01Shepard:1968:TIF:800186.810616}, Kriging \citep{02Krige1951JCMMS}, Discrete 
Smoothing Interpolation (DSI) \citep{03DBLP:journals/tog/Mallet89,04DBLP:journals/cad/Mallet92}, nearest neighbors, 
etc; see a comparative survey investigated by \cite{05DBLP:journals/gandc/FaliveneCTS10}. When applying those 
interpolation algorithms for large-scale datasets, the computational cost is 
in general too high \citep{06DBLP:journals/gandc/HuangY11}. A common and effective solution to 
the above problem is to perform the interpolation procedures in parallel. 
Currently, many efforts have been carried out to parallelize those 
interpolation algorithms in various environments on multi-core CPU and/or 
many-core GPU platforms \citep{07ISI:000320704100005}. 

For example, to accelerate the Kriging method, \cite{08DBLP:journals/gandc/PesquerCP11}
proposed a solution to parallelizing the ordinary Kriging using MPI (Message 
Passing Interface) libraries in an HPC (High Performance Computing) 
environment, and significantly reduced the final execution time of the 
entire process. Similarly, \cite{09DBLP:conf/para/StrzelczykP10} presented a new 
parallel Kriging algorithm to deal with unevenly spaced data. \cite{10DBLP:journals/gandc/Cheng13} proposed an efficient parallel scheme to accelerate 
the universal Kriging algorithm on the NVIDIA CUDA platform by optimizing 
the compute-intensive steps in the Kriging algorithm, such as matrix--vector 
multiplication and matrix--matrix multiplication and achieved a nearly 18 
speedup over the serial program.

 \cite{11DBLP:conf/iccS/AllombertMDBBAJ14} introduced an efficient out-of-core algorithm 
that fully benefited from graphics cards acceleration on a desktop computer, 
and found that it was able to speed up Kriging on the GPU with data 4 times 
larger than a classical in-core GPU algorithm, with a limited loss of 
performances.

To improve the computational efficiency of the most time-consuming steps in 
ordinary Kriging, i.e., the weights calculation and then the estimate for 
each unknown point, \cite{12DBLP:journals/gandc/RaveJAG14} investigated the 
potential reduction in execution time by selecting the suitable operations 
involved in those steps to be parallelized by using general-purpose 
computing on GPUs and CUDA. 

\cite{13DBLP:journals/gandc/HuS15} proposed an improved coarse-grained parallel 
algorithm to accelerate ordinary Kriging interpolation in a homogeneously 
distributed memory system using the MPI (Message Passing Interface) model 
and achieved the speedups of up to 20.8. \cite{14doi:10.1080/15481603.2014.1002379} 
proposed an algorithm based on the $k$-d tree method to partition a big dataset 
into workload-balanced child data groups, and achieved high efficiency when 
the datasets were divided into an optimal number of child data groups.

The IDW interpolation algorithm has been also parallelized on various 
platforms. For example, exemplified by a hybrid IDW algorithm to generate 
DEM from LiDAR point clouds, \cite{15DBLP:journals/gandc/GuanW10} designed and 
implemented a parallel pipeline algorithm for multi-core platforms, and 
processed nearly one billion LiDAR points in about 12 min and produced a 
27,500 $\times $ 30,500 raster DEM using less than 800 MB of main memory on a 
2.0 GHz Quad-Core Intel Xeon platform. \cite{16,17} accelerated the 
IDW method on the GPU for predicting the snow cover depth at the desired 
point.

\cite{18,19DBLP:journals/jzusc/XiaKL11} proposed a generic methodological framework for geospatial 
analysis based on GPU and explored how to map the inherent parallelism 
degrees of IDW interpolation, which gave rise to a high computational 
throughput. \cite{20DBLP:journals/gandc/HuangLTWCH11} explored of the implementation of a 
parallel IDW interpolation algorithm in a Linux cluster-based parallel GIS. 
\cite{21} developed their IDW interpolation application uses the 
Java Virtual Machine (JVM) for the multi-threading functionality.

\cite{22ISI:000332539900001} developed two GPU implementations of the standard IDW 
interpolation algorithm, the tiled version and the CDP version, by 
exploiting the shared memory and CUDA Dynamic Parallelism, and observed that 
the tiled version can achieve the speedups of 120 and 670 over the CPU 
version when the power parameter was set to 2 and 3.0, respectively. \cite{23} 
also evaluated the impact of several data layouts on the efficiency of 
GPU-accelerated IDW interpolation.

Some of the other efforts have been also carried out to parallelize other 
interpolation algorithms. For example, \cite{24DBLP:journals/gandc/WangGY10} presented a 
computing scheme to speed up the Projection-Onto-Convex-Sets (POCS) 
interpolation for 3D irregular seismic data with GPUs. \cite{25DBLP:journals/gis/GuanKG11}
developed a parallel the fast Fourier transform (FFT) based 
geostatistical areal interpolation algorithm in a homogeneously distributed 
memory system using the MPI programming model. \cite{26DBLP:journals/esi/HuangCCLW12}
employed the $k$-d tree in nearest neighbors search to accelerate the grid 
interpolation on the GPU. \cite{27DBLP:conf/fedcsis/CuomoGGS13} proposed a parallel method 
based on radial basis functions for surface reconstruction on GPU. 

The Adaptive IDW (AIDW) is an improved version of the standard IDW \citep{01Shepard:1968:TIF:800186.810616}, which was originally proposed by \cite{28DBLP:journals/gandc/LuW08}. 
The basic and most interesting idea behind the AIDW is that: it attempts to 
calculate the power parameter adaptively according to the spatial 
distribution pattern of the data points, while in the standard IDW the power 
parameter is a user-specified constant value. Due to the adaptive 
determination of the power parameter, the AIDW method can achieve much more 
accurate prediction results than those by the standard IDW. 

In our previous work \citep{29DBLP:journals/corr/MeiXX15}, we have designed and implemented a parallel AIDW 
algorithm on a GPU. And we have also evaluated the performance of the 
parallel AIDW method by comparing its efficiency with that of the 
corresponding serial one. We have observed that our GPU-accelerated AIDW 
algorithm can achieve the speedups of up to 400 for one million data points 
and interpolated points on single precision.

In our previous GPU implementations of the parallel AIDW method, we have 
found that the most computationally intensive step is the $k$ Nearest Neighbors 
($k$NN) search for each interpolated points. We have designed a straightforward 
method to find the $k$ nearest neighboring data points for each interpolated 
point within a single thread. Although the GPU implementing using our 
straightforward $k$NN search approach can achieve satisfied computational 
efficiency, for example, the obtained speedups are about 100 $\sim $ 400 on 
single precision, further performance improvement probably can be achieved by 
optimizing the $k$NN search. 

The task of the $k$NN search is to find the nearest neighbors to an input 
query. Previous research works on the $k$NN search are mainly implemented and 
optimized in CPU \citep{n01DBLP:journals/cg/SankaranarayananSV07}. Recently, GPU-accelerated implementations 
have improved performance by utilizing the massively parallel architecture 
of a single GPU \citep{n02DBLP:conf/cvpr/GarciaDB08,n03DBLP:journals/ijpp/LeiteTFRTK12,n04DBLP:conf/icde/PanM12,n05Liang5382329,06DBLP:journals/gandc/HuangY11, n07DBLP:journals/prl/BeliakovL12, n08DBLP:journals/corr/KomarovDD13, n09DBLP:journals/prl/LiuW15}, multi-GPUs \citep{n10DBLP:journals/concurrency/KatoH12,n11arefin2012gpu}, and GPU clusters \citep{n12}. 
Among those GPU-accelerated $k$NN search algorithms, most of them focusing on 
speeding up the brute-force $k$NN search algorithm; and several of them are 
designed and optimized using space partitioning data structures such as grid 
\citep{n03DBLP:journals/ijpp/LeiteTFRTK12}, RP-tree \citep{n04DBLP:conf/icde/PanM12}, VP-tree 
\citep{n09DBLP:journals/prl/LiuW15}, and $k$-d tree \citep{n07DBLP:journals/prl/BeliakovL12}. 

In this paper, we attempt to improve the efficiency of our previous 
GPU-accelerated AIDW algorithm by adopting a more efficient $k$NN search 
approach. The efficient $k$NN search is expected to be performed in a separate 
stage with the use of the data structure, grid. The resulting values of the 
$k$NN search are the distances between the $k$ nearest neighboring data points to 
each interpolated point. Those distances are then transferred into another 
stage of the AIDW to adaptively calculate the power parameter and the 
expected prediction value (i.e., the weighted average). To evaluate the 
improved parallel AIDW algorithm, we also compare its efficiency with that 
of our previous one introduced in \cite{29DBLP:journals/corr/MeiXX15}.

The rest of this paper is organized as follows. Section 2 introduces the 
background principles of the IDW algorithm, the AIDW algorithm, and the 
$k$NN search. Section 3 describes the strategies and considerations for 
improving our previous GPU-accelerated AIDW algorithm. Section 4 presents 
some implementation details of the improved algorithm. Some comparative 
experimental tests and analysis are provided in Section 5. Finally, Section 
6 draws several conclusions.

\section{Background}
This section will briefly introduce the principles of the standard IDW 
interpolation method \citep{01Shepard:1968:TIF:800186.810616}, the AIDW interpolation method 
\citep{28DBLP:journals/gandc/LuW08}, and the $k$NN search.

\subsection{The Standard IDW Interpolation}
The IDW algorithm is one of the most popular spatial interpolation methods 
in Geosciences, which calculates the prediction values of 
unknown/interpolated points by weighting average of the values of known/data 
points. The name given to this type of methods was motivated by the weighted 
average applied since it resorts to the inverse of the distance to each 
known point when calculating the weights. The difference between different 
forms of IDW interpolation is that they calculate the weights variously. 

A general form of predicting an interpolated value $Z$ at a given point $x$ based 
on samples $Z_{i}=Z(x_{i})$ for $i$ = 1, 2, {\ldots}, $n$ using IDW is an 
interpolating function: 

\begin{equation}
	\label{eq1}
	Z(x)=\sum\limits_{i=1}^n {\frac{\omega _i (x)z_i }{\sum\limits_{j=1}^n 
			{\omega _j (x)} }} ,
	\quad
	\omega _i (x)=\frac{1}{d(x,x_i )^\alpha }.
\end{equation}

The above equation is a simple IDW weighting function, as defined by \cite{01Shepard:1968:TIF:800186.810616}, where $x$ denotes a prediction location, $x_{i }$ is a data 
point, $d$ is the distance from the known data point $x_{i}$ to the unknown 
interpolated point $x$, $n$ is the total number of data points used in 
interpolating, and $p$ is an arbitrary positive real number called the power 
parameter or the distance-decay parameter (typically, $\alpha $ = 2 in the 
standard IDW). Note that in the standard IDW, the power/distance-decay 
parameter $\alpha $ is a user-specified constant value for all unknown 
interpolated points. 

\subsection{The AIDW Interpolation}
The AIDW is an improved version of the standard IDW \citep{01Shepard:1968:TIF:800186.810616}, which is 
originated by \cite{28DBLP:journals/gandc/LuW08}. The basic and most interesting 
idea behind the AIDW is that: it adaptively determines the distance-decay 
parameter $\alpha $ according to the spatial pattern of data points in the 
neighborhood of the interpolated points. In other words, the distance-decay 
parameter $\alpha $ is no longer a pre-specified constant value but 
adaptively adjusted for a specific unknown interpolated point according to 
the distribution of the nearest neighboring data points.

When predicting the desired values for the interpolated points using AIDW, 
there are typically two phases: the first one is to determine adaptively the 
power parameter $\alpha $ according to the spatial pattern of data points; 
and the second is to perform the weighting average of the values of data 
points. The second phase is the same as that in the standard IDW; see 
Equation (\ref{eq1}).

In AIDW, for each interpolated point, the
parameter $\alpha $ can be adaptively determined according to the following steps.

\textbf{Step 1}: Determine the spatial pattern by comparing the observed 
average nearest neighbor distance with the expected nearest neighbor 
distance.

\begin{enumerate}[1)]
	\item Calculate the expected nearest neighbor distance $r_{\exp } $ for a random pattern using:
\begin{equation}
	\label{eq2}
	r_{\exp } =\frac{1}{2\sqrt {n \mathord{\left/ {\vphantom {n A}} \right. 
				\kern-\nulldelimiterspace} A} },
\end{equation}
where $n$ is the number of points in the study area, and $A$ is the area of the 
study region.

	\item Calculate the observed average nearest neighbor distance $r_{obs} $ by taking the average of the nearest neighbor distances for all points:

\begin{equation}
	\label{eq3}
	r_{obs} =\frac{1}{k}\sum\limits_{i=1}^k {d_i } ,
\end{equation}
where $k$ is the number of nearest neighbor points, and $d_i $ is the 
nearest neighbor distances. The $k$ can be specified before interpolating.

	\item Obtain the nearest neighbor statistic $R\left( {S_0 } \right)$ by:

\begin{equation}
	\label{eq4}
	R\left( {S_0 } \right)=\frac{r_{obs} }{r_{\exp } },
\end{equation}
where $S_{0 }$ is the location of an interpolated point.
\end{enumerate}

\textbf{Step 2}: Normalize the $R\left( {S_0 } \right)$ measure to $\mu _R $ 
such that $\mu _R $ is bounded by 0 and 1 by a fuzzy membership function: 
\begin{equation}
\label{eq5}
\mu _R =\left\{ {\begin{array}{ll}
	0&R\left( {S_0 } \right)\le \mbox{ }R_{\min } \mbox{ } \\ 
	0.5-0.5\cos \left[ {\frac{\pi }{R_{\max } }\left( {R\left( {S_0 } 
			\right)-R_{\min } } \right)} \right]&R_{\min } \le R\left( {S_0 } 
	\right)\le \mbox{ }R_{\max } \mbox{ } \\ 
	1&R\left( {S_0 } \right)\ge \mbox{ }R_{\max } \\ 
	\end{array}} \right.,
\end{equation}
	where $R_{\min } \mbox{ }$ or $R_{\max } $ refers to a local nearest neighbor 
	statistic value (in general, the $R_{\min } \mbox{ }$ and $R_{\max } $ can 
	be set to 0.0 and 2.0, respectively).
	
	\textbf{Step 3}: Determine the distance-decay parameter $\alpha $ by mapping 
	the $\mu _{R}$ value to a range of $\alpha _{ }$ by a triangular 
	membership function that belongs to certain levels or categories of 
	distance-decay value; see Equation (\ref{eq6}).
\begin{equation}
\label{eq6}
\alpha \left( {\mu _R } \right)=\left\{ {{\begin{array}{ll}
		{\alpha _1 } & {\mbox{0.0}\le \mu _R \le \mbox{0.1}} \\
		{\alpha _1 \left[ {1-5\left( {\mu _R -0\mbox{.}1} \right)} \right]+5\alpha 
			_2 \left( {\mu _R -0\mbox{.}1} \right)} & {\mbox{0.1}\le \mu _R \le 
			\mbox{0.3}} \\
		{5\alpha _3 \left( {\mu _R -0\mbox{.}3} \right)+\alpha _2 \left[ {1-5\left( 
				{\mu _R -0\mbox{.}3} \right)} \right]} & {\mbox{0.3}\le \mu _R \le 
			0\mbox{.}5} \\
		{\alpha _3 \left[ {1-5\left( {\mu _R -0\mbox{.5}} \right)} \right]+5\alpha 
			_4 \left( {\mu _R -0\mbox{.}5} \right)} & {\mbox{0.5}\le \mu _R \le 
			\mbox{0.7}} \\
		{5\alpha _5 \left( {\mu _R -0\mbox{.7}} \right)+\alpha _4 \left[ {1-5\left( 
				{\mu _R -0\mbox{.7}} \right)} \right]} & {\mbox{0.7}\le \mu _R \le 
			\mbox{0.9}} \\
		{\alpha _5 } & {\mbox{0.9}\le \mu _R \le \mbox{1.0}} \\
		\end{array} }} \right.,
\end{equation}
			where the $\alpha _{1}$, $\alpha _{2}$, $\alpha _{3}$, $\alpha 
			_{4}$, $\alpha _{5}$ are the assigned to be five levels or categories of 
			distance-decay value.
			
			After determining the parameter $\alpha $, the desired prediction value of 
			each interpolated point can be obtained via the weighting average. This 
			stage is the same as that in the standard IDW; see Equation (\ref{eq1}).
			
			\subsection{$k$NN Search}
			The principle and major steps of the brute-force $k$NN search are as follows \citep{n02DBLP:conf/cvpr/GarciaDB08}:
			
			Considering a set $R$ of $m$ reference points in a \textit{d}-dimensional 
			space $R=\{r_1 ,r_2 ,.....,r_m \}$, and a set $Q$ of $n$ query points in the 
			same space $Q = \mbox{\{}q_1 ,q_2 , ... , q_n \mbox{\}}$, for a query point 
			$q \in Q$, the brute-force algorithm is composed of the following steps:
			
			1) Compute the distance between $q$ and the $m$ reference points of $R$:
			
			2) Sort the $m$ distances;
			
			3) Output the distances in the ordered of increasing distance.
			
			When applying this algorithm for the $n$ query points with considering the typical 
			case of large sets, the complexity of this algorithm is overwhelming:
			
			\begin{itemize}
				\item $O(nmd)$ multiplications for the $n\times m$ distances computed;
				\item $O(nm\log m)$ is for the $n$ sorting processes.
			\end{itemize}
			
			The brute-force $k$NN search method is by nature highly parallelizable and 
			perfectly suitable for a GPU implementation.
			
			\section{The Improved GPU-accelerated AIDW Method}
			This section will briefly introduce the considerations and strategies in the 
			development of the improved GPU-accelerated AIDW interpolation algorithm.
			
			\subsection{Overview and Basic Ideas}
			The basic and most interesting concept behind the AIDW method is that: it 
			attempts to determine adaptively the power parameter according to the 
			spatial distribution pattern of each interpolated point. In AIDW algorithm, 
			the spatial distribution pattern is considered as the distribution density 
			of several nearest neighboring data points locating around an interpolated 
			point, which can be roughly measured by using the average distance from 
			those neighboring data points to the interpolated point.
			
			In our previous work, we present a straightforward, easy-to-implement, and 
			suitable for GPU-parallelization algorithm to find the $k$ nearest neighboring 
			data points of each interpolated point. Assuming there are $n$ interpolated 
			points and $m$ data points, for each interpolated point we carry out the 
			following steps \citep{29DBLP:journals/corr/MeiXX15}:
			
			Step 1: Calculate the first $k$ distances between the first $k$ data points and 
			the interpolated points; 
			
			Step 2: Sort the first $k$ distances in ascending order; 
			
			Step 3: For each of the rest ($m-k)$ data points, 
			
			1) Calculate the distance \textit{dist};
			
			2) Compare the \textit{dist} with the $k$th distance:
			
			if \textit{dist} $<$ the $k$th distance, then replace the $k$th distance with the \textit{dist}
			
			3) Iteratively compare and swap the neighboring two distances from the $k$th 
			distance to the first distance until all the $k$ distances are newly sorted in 
			ascending order.
			
			The major advantage of the above algorithm is that: it is simple and easy to 
			implement. Obviously, there is no need to utilize any complex space 
			partitioning data structures such as various types of \textit{trees}. In contrast, only 
			arrays for storing distances and coordinates are needed. Also, we find the 
			desired nearest neighbors without the use of explicit sorting algorithms 
			such as binary search. In general, most sorting algorithms are 
			computationally complex and not suitable for entirely being invoked within a 
			single GPU thread. 
			
			The most obvious shortcoming of the above algorithm for finding nearest 
			neighboring data points is that: it is computationally inefficient due to 
			the global search for nearest neighbors. In that algorithm, the first $k$ 
			distances are calculated and recorded; and then the distances to the rest 
			points are calculated and then compared with those first $k$ distances. The 
			above procedure obviously needs a global search, which is not 
			computationally optimal. One of the frequently used optimization strategies 
			is to perform a local search by filtering those data points and distances 
			that are not needed to be considered. 
			
			In this work, we focus on improving our previous GPU-accelerated AIDW 
			algorithm by using a fast $k$NN search algorithm. Our considerations and basic 
			ideas behind developing the efficient $k$NN search algorithm are as follows: 
			
			(1) Create an even grid to partition the planar region that encloses the 
			projected positions of all data points and interpolated points;
			
			(2) Distribute all the data points and interpolated points into the grid and 
			record the locations;
			
			(3) Perform a \textit{local} and fast search within the grid to find the nearest 
			neighboring data points for each interpolated point.
			
			After obtaining the average distance of those neighboring data points, the 
			adaptive power parameter $\alpha $ will be determined according to the 
			average distance. Finally, the desired prediction value for each 
			interpolated point can be obtained via weighting average using the parameter 
			$\alpha $; see more descriptions in Subsection 2.2.
			
\begin{figure}[ht]
				\centering
				\includegraphics[width=\linewidth]{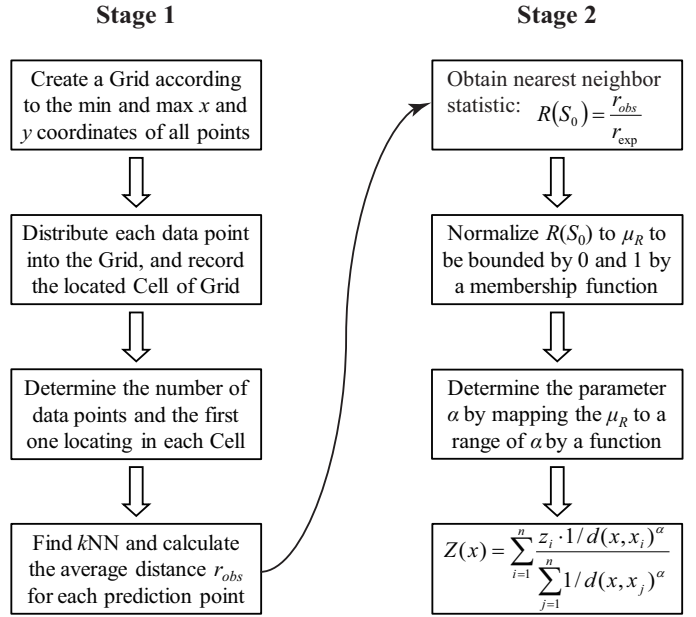}
				\caption{Process of the improved GPU-accelerated AIDW interpolation algorithm}
				\label{fig1}
			\end{figure}

			In summary, the improved GPU-accelerated AIDW algorithm is mainly composed of
			two stages: (1) the $k$NN search and average distances calculation, and (2) the 
			determination of adaptive power parameter and prediction value by weighted 
			interpolating; see Figure \ref{fig1}.
			
			\subsection{Stage 1: $k$NN Search}
			The workflow of the stage of $k$NN search is listed in Figure \ref{fig1}. In this 
			section, more descriptions on this stage will be presented.
			
			\subsubsection{Creating an Even Grid}
			
\begin{figure}[ht]
							\centering
							\includegraphics[width=\linewidth]{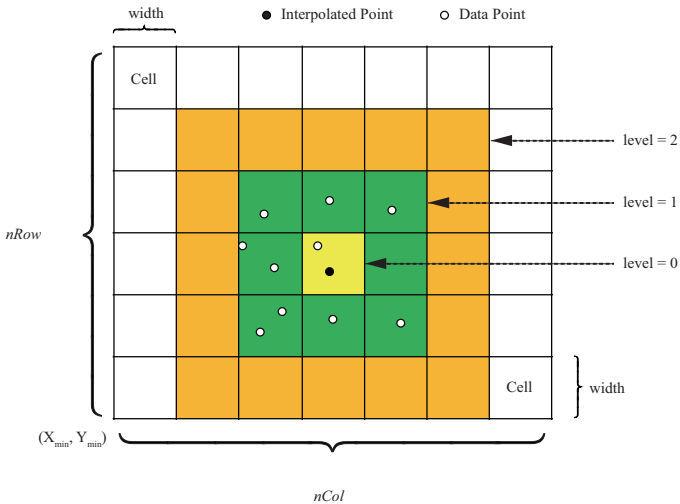}
							\caption{The creation of an even grid according to the minimum and maximum coordinates of all the data points and interpolated points}
							\label{fig2}
						\end{figure}

			The even grid is a simple type of data structure for space partitioning, 
			which is composed of regular cells such as squares or cubes; see an example of 
			planar grid illustrated in Figure \ref{fig2}. Compared to other efficient but complex 
			space partitioning data structures such as the $k$-d tree, the even grid is 
			much easier to create and search objects. In this work, we use a planar even 
			grid to partition all data points to speed up the $k$NN search via local 
			search.
			
			The building of an even planar grid is straightforward. We first calculate 
			or specify the width of the square cell, then determine the planar 
			rectangular region for partitioning according to the minimum and maximum $x$ 
			and $y$ coordinates of all points, i.e., obtain the length and width of the 
			rectangle. After that, the numbers of rows and columns of the grid can be 
			quite easily determined by dividing the rectangle.
			
			\subsubsection{Distributing Data Points into Cells}
			
			The distribution of each data point is to find out that in which grid cell 
			the data point locates. Since each grid cell can be located and recorded 
			using its row and column indices, the distribution of each data point is in 
			fact to obtained the row and column indices of the cell in which it locates. 
			
			This procedure can also be quite easily performed. First, the differences 
			between the coordinates of the data points and the minimum coordinates of 
			all cells are calculated; then the indices of row and column can be obtained 
			by dividing the above differences with the cell width.
			
			\subsubsection{Determining Data Points in Each Cell}
			
			The most important and basic idea behind utilizing a space partitioning is 
			to perform a local search within local regions rather than a global search. 
			When searching nearest neighbors, it is computationally optimal to first 
			search approximate nearest neighbors within several local cells and then to 
			find the exact nearest neighbors by filtering undesired points. 
			
			Since the local search is operated within cells, it is thus needed to 
			determine that which data points locate inside a specific cell. In other 
			words, it is needed to know the number and the indices of those data points 
			locating in the same cell. Moreover, the layout for storing the number and 
			indices should be carefully handled.
			
			For each grid cell, to store the above-mentioned number and indices of those 
			data points locating in the same cell, in general, a dynamic array of 
			integers needs to be allocated. In the traditional CPU computing, the 
			allocation and operations of dynamic arrays are easy-to-implement and 
			computationally inexpensive. However, in GPU computing, it is no longer easy 
			to implement or computationally cheap. This is because that: (1) in GPU 
			computing the programming model such as CUDA cannot support the allocation 
			and operations of dynamic arrays/containers like \texttt{vector} and \texttt{list} in C++ STL 
			(Standard Template Library); and (2) the allocation of a large-enough static 
			array of integers, e.g., \texttt{int index[1000]}, for storing the indices of data 
			points within each GPU thread is not memory efficient. 
			
			Due to the above reasons, we design an optimal layout for storing the number 
			and indices of data points. The basic idea is that: if the indices of those 
			points locating inside the same cell are stored in a continuous 
			segment/piece of integer values, then we only need to know the address of 
			the first point in the segment and the number of points in the same segment 
			(i.e., the size of the segment).
			
			In this case, for each cell, we can only use two integer values to record 
			the number and the indices of those data points that locate in the same 
			cell. One integer is used to hold the number, and the other is used to 
			record the address of the head/first point in each segment. The above two 
			values can be very efficiently determined in a parallel fashion.
			
			\begin{figure}[ht]
				\centering
				\includegraphics[width=\linewidth]{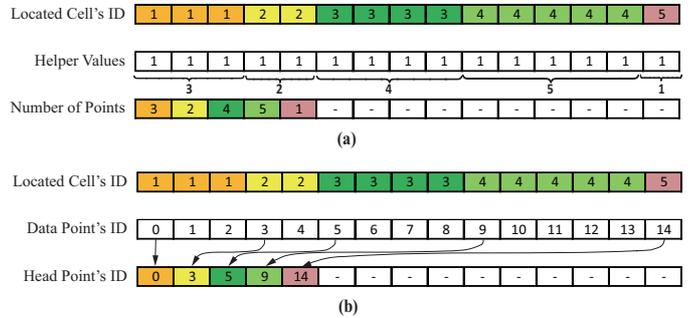}
				\caption{Demonstration of determining the number of data points 
							distributed in each cell and the index of the head point. (a) the number of 
							points; (b) the index of the head point}
				\label{fig3}
			\end{figure}

			Before determining the number and indices of data points locating in the 
			same cell, those data points should be recorded continuously. Since we have 
			obtained the index of the cell in which each data point locates, if we sort 
			all data points according to their corresponding cell indices in ascending 
			order, then those data points locating in the same cell can be gathered in a 
			continuous segment. This sorting procedure is suited to be parallelized on 
			the GPU.
			
			The number of data points locating in the same cell is determined using 
			\textit{segmented} parallel reduction. As described above, after sorting all data points 
			according to cell indices, all data points are stored in a group of 
			segments; each segment is flagged with the cell index, and contains the 
			indices of data points locating in the same cell. The number of data points 
			locating in the same cell can be achieved by performing a reduction for each 
			segment; see Figure \ref{fig3}(a). Similarly, the head index of the first point of 
			each segment can be obtained using segmented parallel scan; see Figure \ref{fig3}(b).
			
			\subsubsection{Searching Nearest Neighbors}
			
			In this work, a space-partitioning data structure, the even grid, is 
			employed to enhance the $k$NN search algorithm. The most important and basic 
			idea behind utilizing the space partitioning is to perform a local search 
			within local regions rather than a global search. This idea is quite 
			effective in practice for that the number of points that are needed to find 
			and compare can be significantly reduced, and therefore, the computational 
			efficiency can be improved.
			
			The process of $k$NN search for each interpolated point can be summarized as 
			follows.
			\begin{itemize}
				\item Step 1: Locate the interpolate point into the even grid
				\item Step 2: Determine the level of cell expanding
				\item Step 3: Find the nearest neighbors within the local region
				\item Step 4: Calculate the average distance
			\end{itemize}
			
			The locating of each interpolated point into the previously created planar 
			grid is quite straightforward. Since each grid cell can be located and 
			recorded using its row and column indices, the distribution of each 
			interpolated point is in fact to obtained the row and column indices of the 
			cell in which it locates. First, the differences between the coordinates of 
			the interpolated point and the minimum coordinates of all cells are 
			calculated; then the indices of row and column can be obtained by dividing 
			the above differences with the cell width.
			
			The determining of the level of cell expanding is in fact to determine the 
			region of cells in which the local nearest neighbors search should be 
			carried out; see three levels of cell expanding in Figure \ref{fig2}. In $k$NN search, 
			the number of nearest neighbors, $k$, is typically pre-specified; and 
			obviously, the number of data points locating in the local cells must be 
			larger than the number $k$. Thus, the level of cell expanding can be 
			iteratively determined by comparing the number of currently found data 
			points with the number $k$. For example, when the $k$ is specified as 15, and 
			within the first level of local cells there are only 10 data points, and 
			thus the level 1 needs to expand to level 2. Similarly, if only 14 data 
			points can be found within the second level of local cells, the level needs 
			to be further expanded to 3. This procedure is iteratively repeated until 
			enough data points have been found.
			
\begin{figure}[ht]
				\centering
				\includegraphics[width=0.6\linewidth]{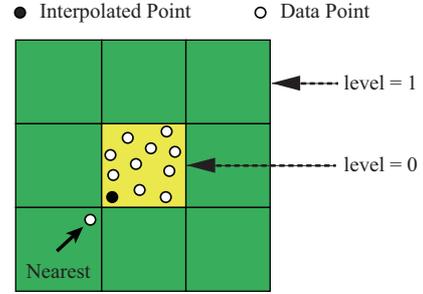}
				\caption{An example for demonstrating the failure of finding exact 
								nearest data points for an interpolated point}
				\label{fig4}
			\end{figure}

			\textbf{Remark:} Note that after iteratively determining the level of cell 
			expanding, for example, level 3, the final level of cell expanding needs to 
			increase with 1, i.e., level 4. This is because that: without expanding 
			additional one level, the nearest neighbors found in the initial level of 
			local cells may not the desired exact $k$ nearest neighbors; see the marked 
			data point in Figure \ref{fig4}. When $k$ = 10, the determined level of cell expanding 
			is 0 (i.e., the yellow region). However, the marked data point is obvious 
			one of the nearest neighbors of the only interpolated point because it is 
			much nearer to the interpolated point than several data points locating in 
			the yellow region. This demonstrates that: without expanding additional one 
			level, incorrect/undesired nearest neighboring data points are probably 
			found; and several of the expected nearest neighboring data points may not able to be found.
			
			The $k$NN search in the local cells is, in fact, to further find exact nearest 
			neighbors by filtering some undesired points. We first allocate an array 
			with the size of $k$ for storing distances, and initiate all distances to 0. 
			Then for each of those data points locating in the local cells, we calculate 
			the distance \textit{dist}, and compare the \textit{dist} with the $k$th distance; and if \textit{dist} is smaller 
			than the $k$th distance, then replace the $k$th distance with the \textit{dist}; after that, we 
			iteratively compare and swap the neighboring two distances from the $k$th 
			distance to the first distance until all the $k$ distances are newly sorted in 
			ascending order; see \cite{29DBLP:journals/corr/MeiXX15} for more details.
			
			After finding the nearest neighbors of each interpolated point, the 
			distances between each nearest neighbor and the interpolated point can be 
			calculated; and finally, the desired average distance can be obtained. 
			
			\subsection{Stage 2: Weighted Interpolating}
			Due to the inherent feature of the AIDW interpolation algorithm, it is 
			perfect that a single GPU thread can take the responsibility to calculate 
			the prediction value of an interpolated point. For example, assuming there 
			are $n$ interpolation points that are needed to be predicted their values such 
			as elevations, and then it is required to allocate $n$ threads to calculate the 
			desired prediction values for all those $n$ interpolated points concurrently. 
			
			In GPU computing, shared memory is inherently much faster than global 
			memory; thus, any opportunity to replace global memory access by shared 
			memory access should therefore be utilized. Since the shared memory 
			residing in the GPU is limited per SM (Stream Multiprocessor), a common 
			optimization strategy called ``tiling'' is frequetnly used to handle the 
			above problem, which partitions the data stored in global memory into 
			subsets called tiles so that each tile fits into the shared memory.
			
			This optimization strategy ``tiling'' is also adopted to accelerate the AIDW 
			interpolation algorithm: the coordinates of all data points are first 
			transferred from global memory to shared memory; then each thread within a 
			thread block can access the coordinates stored in shared memory 
			concurrently. By exploiting the ``tiling'' strategy, the global memory 
			access can be significantly reduced; and thus, the overall computational 
			efficiency is expected to be improved.
			
			\section{Implementation Details}
			As introduced in the above section, the improved GPU-accelerated AIDW 
			interpolation algorithm is mainly composed of two stages, i.e., the $k$NN 
			search stage and the weighted interpolating stage. In this section, we will 
			describe some implementation details on the above two stages. 
			
			\subsection{Stage 1: $k$NN Search}
			\subsubsection{Creating an Even Grid}
			
			An even grid is composed of a group of grid cells, and in this work, each 
			grid cell is a square. The creation of an even grid is in fact to determine 
			the position of the grid, the size of the cell, and the distribution layout 
			of the cells. In our algorithm, an even planar grid is created to cover the 
			planar region in which the projected positions of all data points and 
			interpolated points locate. 
			
			We first obtain the minimum and maximum coordinates of all the data points 
			and interpolated points using the parallel reduction 
			\texttt{thrust::minmax{\_}element()} provided by the library \textit{Thrust} \citep{30Bell2012359}, and 
			calculate the differences between those minimum and maximum coordinates in 
			$x$- and $y$- direction. After approximately determining the planar region, we 
			then calculate the length of interval \texttt{cellWidth}, i.e., the width of a square 
			cell, according to Equation (\ref{eq2}). After that, the number of rows and columns 
			of grid cells can be easily calculated as follows:
			
			\texttt{int nCol = (maxX - minX + cellWidth) / cellWidth;}
			
			\texttt{int nRow = (maxY - minY + cellWidth) / cellWidth;}
			
			\subsubsection{Distributing Data Points into Cells}
			
			After creating the even grid, the subsequent step is to distribute all the 
			data points into the grid. This procedure can be naturally parallelized 
			since the distributing of each data point can be performed independently. 
			Assuming there are $m$ data points, we allocate $m$ GPU threads to distribute all 
			the data points. Each thread is responsible for calculating the position of 
			one data point locating in the grid, i.e., to determine the index of the 
			cell where the data point locates. This can be very easily achieved using 
			the following formulations.
			
			\texttt{int col{\_}idx = (int) (dx[i] - minX) / cellWidth;}
			
			\texttt{int row{\_}idx = (int) (dy[i] - minY) / cellWidth;}
			
			A cell in a grid can be exactly positioned according to the indices of row 
			and column, i.e., \texttt{int col{\_}idx}, \texttt{row{\_}idx}. Also, the position of each 
			grid can be found according to its global index that can be calculated using 
			the simple transformation, \texttt{global{\_}idx = row{\_}idx * nCol + col{\_}idx}. 
			
			The above transformation formulation can be used to transform a 
			two-dimensional index of each grid cell to a unique one-dimensional index. 
			Obviously, this transformation can be easily transformed back. The reason 
			why we carry out the transformation is that: first the memory requirement is 
			reduced since only one array of integers is needed to be stored, and the 
			second is that sorting with using one value as the key is much faster than 
			that with two values as keys. 
			
			To obtain the indices and numbers of those data points locating in each 
			cell, an effective solution is to store those data points that locate in the 
			same cell continuously. Then, operations on the continuous pieces of data 
			(i.e., segments) can be very efficient; see more descriptions in the closely 
			subsequent section.
			
			\subsubsection{Determining Data Points in Each Cell}
			
			In the stage of the $k$NN search, our objective is to find $k$ nearest neighboring 
			data points for each interpolated point. The $k$NN search for each interpolated 
			point is locally performed within several grid cells. The first requirement 
			is to determine how many and which data points locate in each grid cell. 
			More specifically, we need to know the indices and the number of those data 
			points locating in each grid cell. We obtain this simply by using parallel 
			reduction and scan; see our ideas illustrated in Figure \ref{fig3}.
			
			Before carrying out the parallel reduction and scan, those data points that 
			locate inside the same cell should be stored continuously. This requirement 
			can be fulfilled by utilizing a parallel sort with the use of the global index 
			of cells as keys. The parallel sort is realized by using the corresponding 
			parallel primitive provided by the powerful library \textit{Thrust}, 
			\texttt{thrust::sort{\_}by{\_}key(keys, values)}.
			
			Note that those data points locating in the same cell are stored 
			continuously, and if we know the number of data points locating in the same 
			cell, then we only to know the first address of the first data point; and 
			each of the rest data points can be referenced according to the address of 
			the first point and its local position. This idea is quite similar to the 
			reference of any value/element in an array. 
			
			Then, the parallel reduction and scan are also performed by using the 
			primitives provided by \textit{Thrust}. We also use the global index of cells as the keys 
			for \textbf{\textit{Segmented}} reduction and scan. The motivation why we 
			use the segmented reduction and scan rather than the global reduction and 
			scan is that: in the current step we only need to operate on the data points 
			locating in the same cell; and those data points locating in the same cell 
			have been stored continuously and marked using the global index of cell as 
			flags; see Figure \ref{fig3}.
			
			The number of those data points locating in the same cell is obtained by 
			using the primitive \texttt{thrust::reduce{\_}by{\_}keys()}; and the index of the 
			first/head point of each segment of data points are found using 
			\texttt{thrust::unique{\_}by{\_}keys()}. As illustrated in Figure \ref{fig3}, a helper array 
			of constant integers is additionally used to count the number of data points 
			stored in the same piece/segment. 
			
			\subsubsection{Searching Nearest Neighbors}
			
			The finding of $k$ nearest neighboring data points for each interpolated points 
			can be inherently parallelized. Assuming there are $n$ interpolated points, and 
			we allocate $n$ threads to search the nearest neighbors for all the 
			interpolated points. Each thread is invoked to find the nearest neighbors 
			for only one interpolated point. 
			
\begin{figure}[!h]
				\centering
				\includegraphics[width=0.91\linewidth]{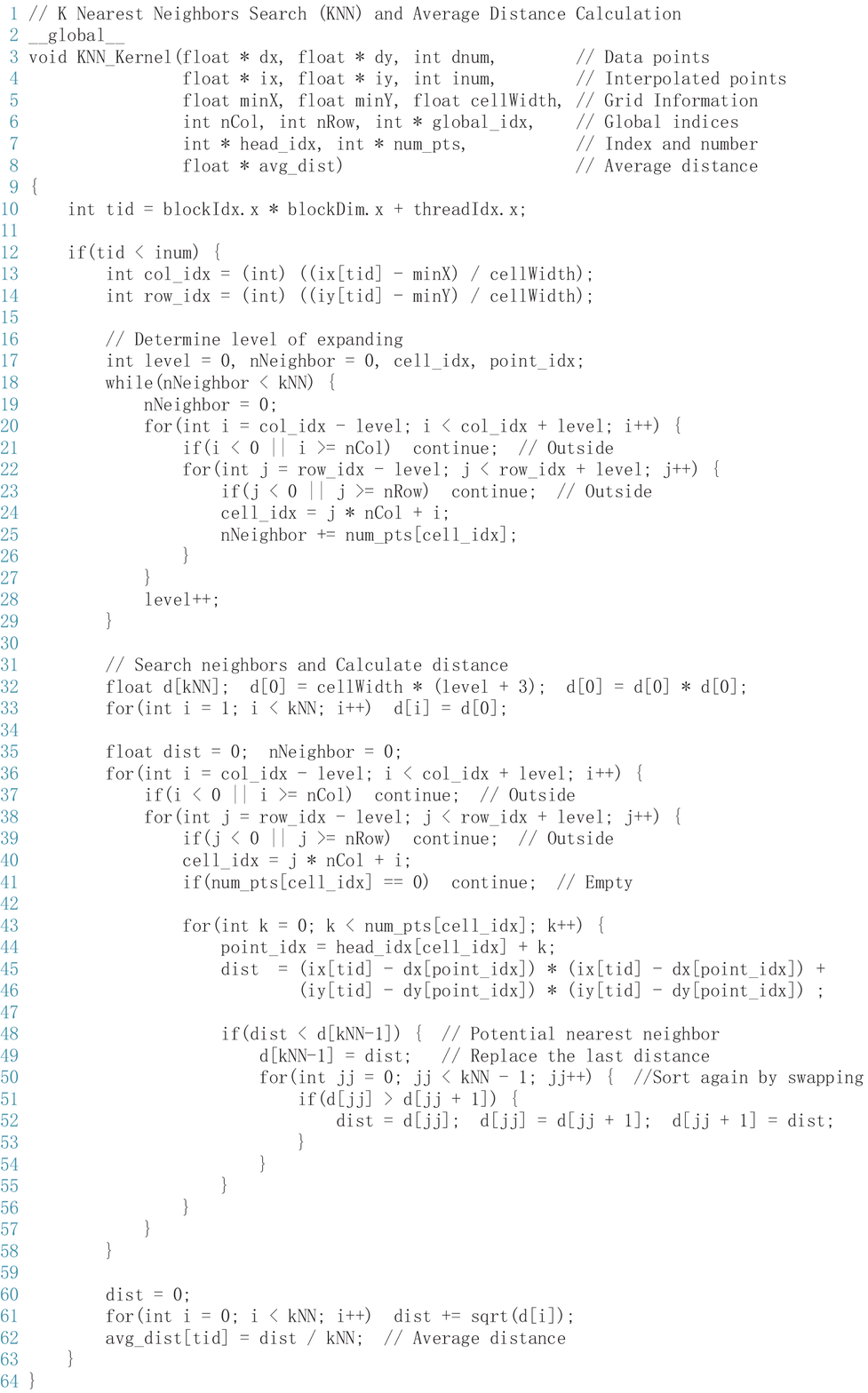}
				\caption{A CUDA kernel of the $k$NN search}
				\label{fig5}
			\end{figure}

			Within each thread, we first distribute the interpolated point into the 
			created grid by calculating its row index and column index; see lines 13 
			$\sim $ 14 in Figure \ref{fig5}. Then we determine the region of the local cells by 
			approximately calculating the level of expanding according to the number of 
			data points; see lines 16 $\sim $ 29 in Figure \ref{fig5}. Note that currently those 
			data points locating in the determined local cells are the 
			\textbf{\textit{Approximate}} nearest neighbors of the interpolated points. 
			After that, we further find the \textbf{\textit{Exact}} nearest neighbors by 
			filtering those approximate nearest neighbors by inserting and swapping; see 
			lines 31 $\sim $ 58 in Figure \ref{fig5}. Finally, the desired average distance 
			between the exact nearest neighboring data points and the target 
			interpolated point is calculated. 
			
			A remarkable implementation detail is that: when finding the nearest 
			neighbors according to the \textit{Euclidean} distances between points, we do not use the real 
			distance value but the square value of the distance. This is because that: 
			in GPU computing the calculation of square root is quite computationally 
			expensive; and any choice to avoid the use of calculating square root should 
			be exploited. Thus, we calculate the square root in the last step of 
			computing the average distance, rather in the step of searching nearest 
			neighbors.

			\subsection{Stage 2: Weighted Interpolating}
			This subsection will present the details on implementing the interpolating 
			stage in the GPU-accelerated AIDW algorithm. We implement two versions: the 
			\textit{naive} version and the \textit{tiled} version, by employing the data layout Structure-of-Arrays 
			(SoA) only. Both the naive and the tiled implementations developed in this 
			work are the same as those corresponding implementations presented in our 
			previous work \citep{29DBLP:journals/corr/MeiXX15}.
			
			\subsubsection{Naive Version}
			
			In this version, the global memory and registers on GPU architecture are 
			employed without exploiting the shared memory. The input data and the output 
			data are stored in the global memory. Assuming that there are $m$ data points 
			used to evaluate the prediction values for $n$ interpolation points, we 
			allocate $n$ threads to parallelize the interpolating. 
			
			The data layout SoA is employed in this version. The coordinates of all data 
			points and interpolated points are stored in the arrays \texttt{float dx[dnum]}, \texttt{dy[dnum]}, \texttt{dz[dnum]}, \texttt{ix[inum]}, \texttt{iy[inum]}, and \texttt{iz[inum]}. 
			
			Since that after invoking the $k$NN kernel, we have obtained the average 
			distance, i.e., the $r_{obs} $ defined in Equation (\ref{eq3}), thus in this stage 
			each thread is only responsible for computing the $r_{\exp } $ and $R\left( 
			{S_0 } \right)$ according to the Equations (\ref{eq2}) and (\ref{eq4}). After that, the 
			$R\left( {S_0 } \right)$ measure is normalized to $\mu _R $ such that $\mu 
			_R $ is bounded by 0 and 1 by a fuzzy membership function; see Equation (\ref{eq5}). 
			Finally, the power parameter $\alpha $ is determined by mapping the $\mu 
			_{R}$ values to a range of $\alpha _{ }$ by a triangular membership 
			function; see Equation (\ref{eq6}).
			
			After adaptively determining the power parameter, the desired prediction 
			value of each interpolated point can be achieved by weighting average. This 
			step of calculating the weighting average is the same as that in the 
			standard IDW method.
			
			\subsubsection{Tiled Version}
			
			The workflow of the tiled version is the same as that of the naive version. 
			The major difference between the two versions is that: in this version, the 
			shared memory is exploited to improve the computational efficiency. 
			
			In the tiled version, the tile size is directly set to be identical to the 
			block size. Each thread within a thread block is invoked to load the 
			coordinates of one data point from global memory to shared memory and then 
			compute the distances and corresponding inverse weights to those data points 
			stored in current shared memory. After all threads within a block finished 
			computing these partial distances and weights, the next piece of data in 
			global memory is loaded into shared memory and used to calculate current 
			wave of partial distances and weights. After calculating each wave of 
			partial distances and weights, each thread accumulates the results of all 
			partial weights and all weighted values into two registers. Finally, the 
			prediction value of each interpolated point can be obtained according to the 
			sums of all partial weights and weighted values and then written into global 
			memory.
			
			By employing the strategy ``tiling'', the global memory access can be 
			significantly reduced for that the coordinates of all data points are only 
			read ($n $/ threadsPerBlock) times rather than $n$ times from global memory, where 
			$n$ is the number of interpolated points and threadsPerBlock denotes the number 
			of threads per block.

			\section{Results and Discussion}
			
			\subsection{Experimental Environment and Testing Data }
			In this work, we focus on improving our previous GPU-accelerated AIDW 
			algorithm by utilizing a fast $k$NN search method. We refer our previously 
			developed GPU-accelerated AIDW algorithm as the \textit{original} algorithm, and the 
			presented algorithm in this work as the \textit{improved} algorithm. 
			
			To evaluate the computational efficiency of the improved algorithm, we have 
			carried out five groups of experimental tests on a laptop computer. The 
			computer is featured with an Intel Core i7 CPU (2.40GHz), 4.0 GB RAM memory, 
			and a GeForce GT730M card. All the experimental tests are executed on OS 
			Windows 7 Professional (64-bit), Visual Studio 2010, and CUDA v7.0.
			
			Two versions of the improved GPU-accelerated AIDW, i.e., the naive version 
			and the tiled version, are implemented using the SoA layout and evaluated on 
			single precision. In contrast, the CPU version of the AIDW implementation is 
			tested on double precision; and all results of this CPU version 
			presented in our previous work \citep{29DBLP:journals/corr/MeiXX15} are directly accepted to be used as the 
			baseline. The efficiency of all GPU implementations is benchmarked by 
			comparing to the baseline results.
			
			When evaluating the execution time of GPU implementations, the overhead 
			spent on transferring the input data (i.e., the coordinates of data points 
			and interpolated points) from the host side to the device side and 
			transferring the results from the device side to the host side is 
			considered. However, the time spent on generating the test data is not 
			included. 
			
			The input of the AIDW interpolation is the coordinates of data points and 
			interpolated points. The efficiency of the CPU and GPU implementations may 
			differ due to different sizes of input data. However, the research objective 
			in this work is to improve our previous GPU-accelerated AIDW algorithm using 
			fast $k$NN search; thus, we only consider a particular situation where the 
			numbers of interpolated points and data points are identical.
			
			All the testing data including the data points and interpolated points are 
			randomly generated within a square. We design five groups of sizes, i.e., 
			10K, 50K, 100K, 500K, and 1000K, where one K represents the number of 1024 
			(1K = 1024). Five tests are performed by setting the numbers of both the 
			data points and interpolated points as the above five groups of sizes.
			
			\subsection{Performance of the Improved GPU-accelerated AIDW Algorithm}
			\subsubsection{Executing Time and Speedups}
			
			We evaluate the computational efficiency of the improved GPU-accelerated 
			AIDW algorithm with the use of five groups of testing data. The running time 
			is listed in Table \ref{tab1}. Note that, to compare with the original 
			GPU-accelerated algorithm, we have also listed the execution time of the 
			original algorithm in Table \ref{tab1}; and these experimental results of the original algorithm are directly 
			derived from our previous work \citep{29DBLP:journals/corr/MeiXX15}.
			
			\begin{table}[htbp]
				\caption{Execution time (/ms) of CPU and GPU versions of the AIDW 
					algorithm on single precision}
				\begin{center}
					\small
					\begin{tabular}{p{50pt}p{20pt}p{20pt}p{30pt}p{30pt}p{30pt}}
        \toprule
						\raisebox{-1.50ex}[0cm][0cm]{Version}& 
						\multicolumn{5}{c}{Data Size (1K = 1024)}  \\
						\cline{2-6} 
						& 
						10K& 
						50K& 
						100K& 
						500K& 
						1000K \\
			\midrule
						CPU/Serial& 
						6791& 
						168234& 
						673806& 
						16852984& 
						67471402 \\
						Original naive version& 
						65.3& 
						863& 
						2884& 
						63599& 
						250574 \\
						Original tiled version& 
						61.3& 
						714& 
						2242& 
						43843& 
						168189 \\
						Improved naive version& 
						27.9& 
						400& 
						1366& 
						31306& 
						124353 \\
						Improved tiled version& 
						21.0& 
						233& 
						771& 
						16797& 
						66338 \\
			\bottomrule
					\end{tabular}
					\label{tab1}
				\end{center}
			\end{table}
			
			\begin{figure}[!h]
				\centering
				\includegraphics[width=0.9\linewidth]{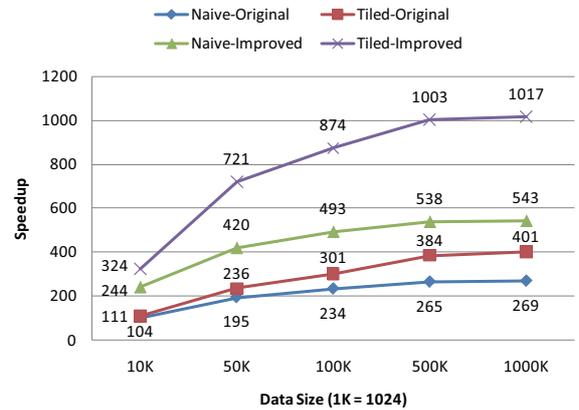}
				\caption{Speedups of the improved and the original GPU-accelerated 
								AIDW algorithms over the serial AIDW algorithm}
				\label{fig6}
			\end{figure}
			
			We have also calculated the speedups of our improved GPU-accelerated AIDW 
			algorithm against the corresponding serial algorithm (i.e., the CPU version 
			listed in Table \ref{tab1}); see Figure \ref{fig6}. The results indicate that: (1) the highest 
			speedups achieved by the naive version and the tiled version can be up to 
			543 and 1017, respectively; and (2) the tiled version is always faster than 
			the naive version.
			
			\subsubsection{Comparison of the Improved Naive Version and Tiled Version}
			
			As observed from the experimental tests, the tiled version of the improved 
			algorithm is about 1.33 $\sim $ 1.87 times faster than the naive version. 
			This behavior is due to the reason that: the stage of interpolating in the 
			tiled version is much more computationally efficient than that in the naive 
			version; see the execution time of the interpolating stage in Table \ref{tab2}.
			
			As described in Section 3, the improved algorithm includes both the naive 
			version and tiled version, which can be divided into two major stages: i.e., 
			the stage of $k$NN search and the stage of weighted interpolating. The first 
			stage in the above two versions are the same, while the second stage 
			differs. 
			
			In the stage of interpolating of the tiled version, the benefit of the use 
			of shared memory is exploited, while in the naive version it is not. For 
			this reason, the interpolating stage in the tiled version executes about 
			1.79 $\sim $ 1.89 times faster than that in the naive version. Thus, the 
			entire tiled version is more efficient than the naive version. 
			
			\begin{table}[htbp]
			\caption{Execution time (/ms) of the stage of $k$NN search and the 
							stage of weighted interpolating in the improved GPU-accelerated AIDW 
							algorithm}
				\begin{center}
										\small
						\begin{tabular}{p{90pt}p{15pt}p{15pt}p{20pt}p{20pt}p{20pt}}
							\toprule
							\raisebox{-1.50ex}[0cm][0cm]{Stage}& 
							\multicolumn{5}{c}{Data Size (1K = 1024)}  \\
							\cline{2-6} 
							& 
						10K& 
						50K& 
						100K& 
						500K& 
						1000K \\
						\midrule
						$k$NN Search \par (Both versions)& 
						12.3& 
						36& 
						81& 
						440& 
						917 \\
						Weighted Interpolating \par (Improved naive version)& 
						15.6& 
						364& 
						1286& 
						30866& 
						123437 \\
						Weighted Interpolating \par (Improved tiled version)& 
						8.7& 
						197& 
						691& 
						16357& 
						65421 \\
						\bottomrule
					\end{tabular}
					\label{tab2}
				\end{center}
			\end{table}
			
			\subsubsection{Workload between the Stages of \textit{k}NN Search and Weighted Interpolating}
			
\begin{figure}[h!]
				\centering
				\subfigure[Naive version]{
					\label{fig7a}       
									\includegraphics[width=0.85\linewidth]{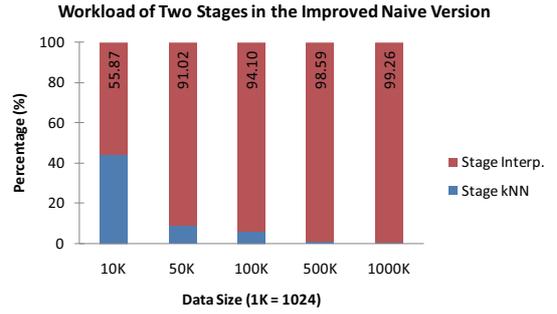}
				}
				\hspace{1em}
				\subfigure[Tiled version]{
					\label{fig7b}       
									\includegraphics[width=0.85\linewidth]{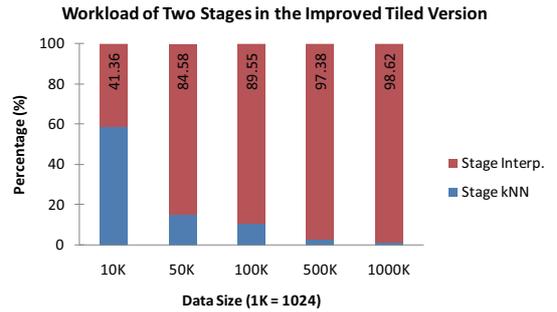}
				}
				\caption{ Workload of the two stages in the improved GPU-accelerated 
					AIDW algorithm}
				\label{fig7}       
			\end{figure}

			There are two major stages in the improved GPU-accelerated AIDW algorithm. 
			To understand the efficiency bottleneck for further optimizations in the 
			future, we in particular record the execution time for the stages of $k$NN 
			search and weighted interpolating separately; see Table \ref{tab2}. In addition, we 
			have also evaluated the workload percentage between the above two stages in 
			both the naive version and tiled version; see Figure \ref{fig7}.
			
			We have found that: the computational cost spent in the stage of $k$NN search 
			is much less than that in the stage of the weighted interpolating. Moreover, 
			with the increase of the size of testing data, the weight of the running 
			time cost in the stage of $k$NN significantly decreases; and it even reduces to 
			about one percentage. This observation indicates that most overhead in both 
			the naive version and the tiled version is spent in the stage of weighted 
			interpolating rather than the $k$NN search. Therefore, further optimizations 
			may need to be employed to improve the efficiency of the weighted 
			interpolating. 
			
			\subsection{Comparison with the Original GPU-accelerated AIDW Algorithm}
			In Section 5.1, we have evaluated the efficiency of the improved algorithm 
			by comparing it with the serial AIDW algorithm, and found that our improved 
			algorithm can achieve quite satisfied speedups. In this section, we will 
			compare our improved GPU-accelerated algorithm presented in this work with 
			the original GPU-accelerated algorithm introduced in \citep{29DBLP:journals/corr/MeiXX15}.
			
			\begin{figure}[!h]
				\centering
				\includegraphics[width = 0.9\linewidth]{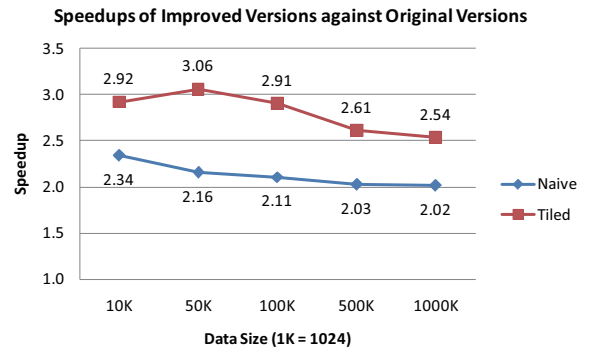}
				\caption{Speedups of the improved GPU-accelerated AIDW algorithm 
					over the original algorithm for both the naive version and tiled version}
				\label{fig8}
			\end{figure}
			
			The speedups of the improved algorithm over the original algorithm are 
			illustrated in Figure \ref{fig8}. The results show that the improved naive version 
			and tiled version are at least 2.02 and 2.54 times faster than the original 
			naive version and tiled version, respectively. This also indicates that 
			significant performance gains have been achieved by improving the original 
			algorithm using fast $k$NN search.
			
			The major difference between the original algorithm and the improved 
			algorithm is the use of different $k$NN search approaches. We attempt to 
			explain the reason why significant performance gains have been achieved by analyzing 
			the impact of different $k$NN search algorithm on the computational efficiency. 
			
			First, we obtain the computational time of the $k$NN search in the original 
			algorithm by subtracting the time spent in the stage of weighted 
			interpolating from the total execution time; see Table \ref{tab3}. Note that, the 
			execution time cost in the stage of weighted interpolating is directly 
			derived from the improved algorithm. This is because that: (1) the weighted 
			interpolating in both the original algorithm and the improved algorithm is 
			the same; and (2) the running time of the weighted interpolating can be 
			separately measured in the improved algorithm, while in contrast it is 
			unable to accurately evaluate the execution time specifically for the 
			weighted interpolating in the original algorithm. 
			
			\begin{table}[htbp]
			\caption{Execution time (/ms) of the stage of $k$NN search in the 
							original and the improved GPU-accelerated AIDW algorithm}
				\begin{center}
					\small
					\begin{tabular}{p{90pt}p{15pt}p{15pt}p{20pt}p{20pt}p{20pt}}
						\toprule
						\raisebox{-1.50ex}[0cm][0cm]{Version}& 
						\multicolumn{5}{c}{Data Size (1K = 1024)}  \\
						\cline{2-6} 
						& 
						10K& 
						50K& 
						100K& 
						500K& 
						1000K \\
						\midrule
						Original naive version& 
						49.7& 
						499& 
						1598& 
						32733& 
						127137 \\
						Original tiled version& 
						52.6& 
						517& 
						1551& 
						27486& 
						102768 \\
						Two improved versions& 
						12.3& 
						36& 
						81& 
						440& 
						917 \\
						\bottomrule
					\end{tabular}
					\label{tab3}
				\end{center}
			\end{table}
			
\begin{figure}[!h]
				\centering
				\includegraphics[width = 0.9\linewidth]{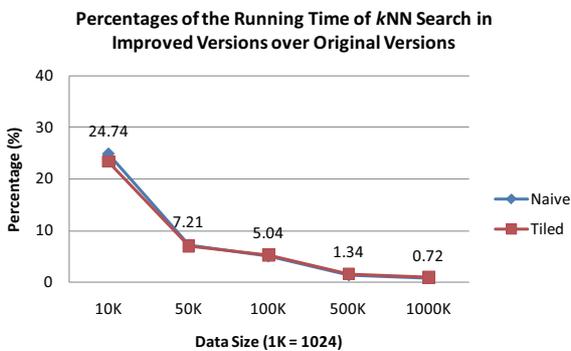}
				\caption{Percentages of the running time of $k$NN search in the 
					improved algorithm over the original algorithm}
				\label{fig9}
			\end{figure}

			Second, we calculate the percentages of the running time of the $k$NN search in 
			the improved algorithm over that in the original algorithm; see Figure \ref{fig9}. We 
			have found that: in both the naive version and the tiled version, the 
			execution time of the $k$NN search in the improved algorithm is much less than 
			that in the original algorithm, for example, less than one percentage for 
			about one million points. This suggests that: the use of fast $k$NN search 
			approach can significantly improve the efficiency of the entire 
			GPU-accelerated AIDW interpolation algorithm. 

			\section{Conclusion}
			In this work, we have presented an efficient AIDW interpolation algorithm on 
			the GPU by utilizing a fast $k$NN search method. The presented algorithm is 
			composed of two major stages, i.e., the $k$NN search and weighted 
			interpolating, and is developed by improving a previous GPU-accelerated AIDW 
			algorithm with the use of fast $k$NN search. The $k$NN search is carried out based 
			upon an even grid, and is capable of finding exact nearest neighbors very 
			fast for each interpolated point. We have performed five groups of 
			experimental tests to evaluate the performance of the improved 
			GPU-accelerated AIDW algorithm. We have found that: (1) the improved 
			algorithm can achieve a speedup of up to 1017 over the corresponding serial 
			algorithm for one million points; (2) the improved algorithm is at least two times faster than our 
			previously developed GPU-accelerated AIDW algorithm; and (3) the utilization 
			of fast $k$NN search can significantly improve the computational efficiency of 
			the entire GPU-accelerated AIDW algorithm. To benefit the community, all 
			source code and testing data related to the presented AIDW algorithm is 
			publicly available.

			\section*{Acknowledgments}
			This research was supported by the Natural Science Foundation of China 
			(Grant No. 40602037 and 40872183), China Postdoctoral Science Foundation 
			(2015M571081), and the Fundamental Research Funds for the Central 
			Universities (2652015065). The authors would like to thank the editor and 
			the reviewers for their contributions on the paper. 

\section*{References}
			
  \bibliographystyle{elsarticle-harv} 
  \bibliography{MG}




\end{document}